\documentclass[11pt]{jpp}
\usepackage{graphicx}
\usepackage{epstopdf, epsfig}
 \usepackage{color}

\newcommand{\be}{\begin{displaymath}}
\newcommand{\bn}{\begin{equation}}
\newcommand{\bea}{\begin{eqnarray*}}
\newcommand{\eea}{\end{eqnarray*}}
\newcommand{\en}{\end{equation}}
\newcommand{\ee}{\end{displaymath}}

\renewcommand{\p}{\partial}
\newcommand{\lang}{\left\langle}
\newcommand{\rang}{\right\rangle}

\shorttitle{Stellarator bootstrap current and plasma flow velocity}
\shortauthor{P. Helander, F.I. Parra and S.L. Newton}

\title{Stellarator bootstrap current and plasma flow velocity at low collisionality}

\author{P. Helander\aff{1}
  \corresp{\email{per.helander@ipp.mpg.de}}
 \and F.I. Parra\aff{2,3}
 \and S.L. Newton\aff{3,4}}

\affiliation{\aff{1}Max-Planck-Institut f\"ur Plasmaphysik, 17491 Greifswald, Germany
\aff{2}Rudolf Peierls Centre for Theoretical Physics, University of Oxford, Oxford OX1 3NP, United Kingdom
\aff{3}Culham Centre for Fusion Energy, Culham, United Kingdom
\aff{4}Department of Physics, Chalmers University of Technology, G\"oteborg, Sweden}

\begin{document}

\maketitle

\begin{abstract}
The bootstrap current and flow velocity of a low-collisionality stellarator plasma are calculated. As far as possible, the analysis is carried out in a uniform way across all low-collisionality regimes in general stellarator geometry, assuming only that the confinement is good enough that the plasma is approximately in local thermodynamic equilibrium. It is found that conventional expressions for the ion flow speed and bootstrap current in the low-collisionality limit are accurate only in the $1/\nu$-collisionality regime and need to be modified in the $\sqrt{\nu}$-regime. The correction due to finite collisionality is also discussed and is found to scale as $\nu^{2/5}$.
\end{abstract}

\section{Introduction}

Calculating the neoclassical transport in stellarators is complicated. Only in the limit of high collisionality, the so-called Pfirsch-Schl\"uter regime, can an accurate analytical calculation be completed using the full, linearised Landau collision operator [\cite{Simakov-2009,Braun-2010}]. Most plasmas of interest are, however, in the opposite regime of low collisionality, which is divided into sub-regimes of different character, where the transport is proportional to $1/\nu$, $\sqrt{\nu}$ or $\nu$, in order of decreasing collisionality, where $\nu$ is the collision frequency. (When the radial electric field is sufficiently small, there can be a so-called superbanana-plateau regime at collisionalities between those where the transport scales as $1/\nu$ and $\nu$, but this regime is only important in particular magnetic geometries and will be ignored in the present paper.) In practice, the various collisionality regimes are usually not well separated from each other, and the true transport must be calculated numerically [\cite{Beidler-2011}]. 

An exception occurs, formally at least, in quasisymmetric and omnigeneous stellarators, where all collisionless orbits are perfectly confined.\footnote{ Quasisymmetric magnetic fields are those in which the field strength is periodic along field lines, $B(l+L) = B(l)$ where $L$ only depends on the magnetic surface. Omnigenous fields are those where the average radial particle drift vanishes. Other definitions of these concepts and how they relate to each other have been given in many places, including \cite{Helander-2014-a}.} In such systems, the problem of calculating the neoclassical transport can be reduced to the corresponding problem in an axisymmetric tokamak. In quasisymmetric systems, this is obvious from the fact that the orbits, and hence the drift kinetic equation, are ``isomorphic'' to those in a tokamak [\cite{Boozer-1983,Landreman-2011}], but even in systems that are merely omnigenous it is, less trivially, possible to split the distribution function into two terms, the first of which can be calculated exactly and the second  of which is determined by the same drift kinetic equation as in a tokamak [\cite{Helander-2009,Landreman-2012}]. It is thus possible to calculate the transport analytically insofar as this is possible in a tokamak, i.e., if the exact collision operator for each species $a$ is replaced by an operator of the form
	\bn C_a(f_{a1}) \rightarrow \nu_a \left( {\cal L} (f_{a1}) + \frac{m_a v_\|}{T_a} w_a f_{a0} \right), 
	\label{C}
	\en
where $\nu_a$ is the collision frequency, 
	$$ {\cal L} = \frac{1}{2} \frac{\p}{\p \xi} \left( 1 - \xi^2 \right) \frac{\p}{\p \xi} $$
the Lorentz operator in terms of $\xi = v_\|/v$, $w_a$ a function of the speed $v$ determined by integral conditions, such as conservation of momentum, and the other symbols have their usual meanings. This operator is not a bad approximation for low collisionality and large aspect ratio, and the problem of calculating the transport in a perfectly omnigenous stellarator can therefore be considered to be solved. The result is, however, of red limited applicability because the deviation from omnigeneity is, in practice, always large enough that the transport has a clear $1/\nu$-regime at low enough collisionality. The deviations from omnigeneity (or quasisymmetry) have to be extremely small in order for tokamak-like levels of neoclassical transport to prevail at low collisionality [\cite{Calvo-2013, Calvo-2014, Calvo-2015, Calvo-2016}]. Even in HSX, the most quasisymmetric of stellarators, there is barely a discernible ``banana'' regime of tokamak-like transport -- see Fig.
\,14 of \cite{Beidler-2011}. 
	
In the present paper, we calculate the flow of plasma within, rather than across, magnetic flux surfaces. Our principal aim is to calculate the bootstrap current in the $\sqrt{\nu}$-regime, which has not been done analytically before. The collisionality is assumed to be small, but the geometry is kept as general as possible, allowing for a $1/\nu$-regime. We also assume that the electrostatic potential is a flux function to the order of interest, and that the plasma is close to a radially local thermodynamic equilibrium, that is, the distribution function is close to a stationary Maxwellian with density and temperature that are flux functions. In general, for very-low-collisionality regimes (collisionalities corresponding to the $\sqrt{\nu}$ and $\nu$ regimes), trapped particles will drift radially distances of the order of the minor radius of the device, making it impossible to achieve a radially local thermodynamic equilibrium [\cite{Helander-2014-a, Calvo-2016}]. We thus need extra conditions to be satisfied to be able to treat very low collisionalities, and are aware of two different assumptions in the literature that lead to the required condition of locally radial thermodynamic equilibrium: closeness to an omnigeneous configuration [\cite{Calvo-2016}], and the condition that the ${\bf E} \times {\bf B}$ drift is much larger than the $\nabla B$ and curvature drifts (valid for large-aspect-ratio stellarators or regions of very large plasma gradients) [\cite{Ho-1987}]. We try to be as general as possible, and we derive equations that are valid for these two possible assumptions.

For a plasma in radially local thermodynamic equilibrium, each species satisfies the leading-order force-balance relation
	$$ n_a e_a ({\bf V}_a \times {\bf B} - \nabla \Phi ) = \nabla p_a, $$
where $n_a$ denotes the density, $e_a$ the charge, ${\bf V}_a$ the flow velocity and $p_a = n_a T_a$ the pressure of the species in question. The magnetic field is in Clebsch form, ${\bf B} = \nabla \psi \times \nabla \alpha$, where $\psi$ labels the magnetic surfaces and $\alpha$ the various field lines on each such surface. The density, temperature and electrostatic potential are to lowest order constant on flux surfaces, which forces the flow velocity to be of the form
	$$ {\bf V}_a(\psi,\alpha, l) = \omega_a \frac{{\bf b} \times \nabla \psi}{B} + V_{a\|} {\bf b}, $$
where ${\bf b} = {\bf B}/B$, $l$ denotes the arc length along the field, and 
	$$ \omega_a(\psi) = \frac{1}{n_a e_a} \frac{dp_a}{d \psi} + \frac{d \Phi}{d \psi}. $$
Incompressibility, $\nabla \cdot {\bf V}_a = 0$, then implies
	$$ V_\|(\psi,\alpha, l) = \omega_a(\psi) \left[ u(\psi,\alpha, l) + k_a(\psi) \right] B, $$
where the function $u(\psi,\alpha, l)$ satisfies the magnetic differential equation
	\bn {\bf b} \cdot \nabla u = - ({\bf b} \times \nabla \psi) \cdot \nabla \left(\frac{1}{B^2} \right), 
	\label{eq for u}
	\en
and $k_a(\psi)$ denotes an arbitrary integration constant. The kinetic theory developed below gives the flux function $k_a(\psi)$.
	
As we shall see, in the $1/\nu$-regime of collisionality the low-collisionality results of 
\cite{Shaing-1983} and \cite{Shaing-1989} are recovered, who derived it using the Hirshman-Sigmar moment method [\cite{Hirshman-1981,Helander-2002-c}]. However, we find that this result ceases to be valid in the $\sqrt{\nu}$ or $\nu$-regimes, where the component of the drifts tangential to the flux surface causes the appearance of an additional term that needs to be taken into account. It is well known from numerical results [\cite{Beidler-2011,Helander-2011-a}] that the approach to the low-collisionality limit is slow, and in the penultimate section of the paper we identify a possible reason for this behaviour and derive a scaling for the deviation from the low-collisionality result. Our conclusions are summarised in the final section. 

\section{Drift kinetic equation}

We begin by dividing the distribution function of each species into even and odd parts in $v_\|  = \sigma |v_\||$,
	$$ f_a({\bf r}, \epsilon, \mu, \sigma ) = f_a^+({\bf r}, \epsilon, \mu) + f_a^-({\bf r}, \epsilon, \mu), $$
	$$ f_a^\pm ({\bf r}, \epsilon, \mu)  = \frac{1}{2} \left[f_a({\bf r}, \epsilon, \mu, +1) \pm f_a({\bf r}, \epsilon, \mu, -1) \right], $$
where $\epsilon = m_a v^2/2 + e_a \Phi({\bf r})$ denotes the (kinetic + potential) energy and $\mu = m_a v_\perp^2/(2B)$ the magnetic moment. The drift kinetic equation is thus split into the two equations
	\bn v_\| \nabla_\| f^- =  C^+(f) - {\bf v}_d \cdot \nabla f^+, \label{+} \en
	\bn v_\| \nabla_\| f^+ = C^-(f)- {\bf v}_d \cdot \nabla f^- , \label{-} \en
where ${\bf v}_d = {\bf v}_E + {\bf v}_M$ is the sum of the ${\bf E} \times {\bf B}$ and magnetic drift velocities and $C^\pm(f)$ denotes the even and odd parts of the collision operator $C(f)$. Thus, for the operator (\ref{C})
	$$ C^+(f_1) = \nu {\cal L} (f^+_{1}) , $$
	$$ C^-(f_1) = \nu \left( {\cal L} (f^-_{1}) + \frac{m v_\|}{T} w f_{0} \right). $$
Here we have omitted the species index, and shall often do so in the following. It is assumed that the electrostatic potential only depends on $\psi$ in leading order -- an assumption that will be discussed below. We can thus set $\Phi = 0$ on the particular flux surface we are considering, but must retain a non-zero derivative $\Phi'(\psi)$. Since $\Phi = 0$, we can use the velocity-space coordinate $v$ instead of $\epsilon$. As usual, particle orbits can be divided into trapped and untrapped (or circulating) ones depending on whether the parameter $\lambda = \mu / \epsilon$ does, or does not, exceed $1/B_{\rm max}$, where $B_{\rm max}(\psi)$ denotes the maximum value of the magnetic field strength on the surface. An important difference between tokamaks and stellarators is that in the latter the field strength generally reaches $B_{\rm max}$ at a discrete set of {\em points} on each surface, whereas $B=B_{\rm max}$ along a {\em circle} on the inboard side of a tokamak. 

The orbit average of an arbitrary function $g(\psi, \alpha, l, \epsilon, \mu, \sigma )$ is defined by 
	$$ \overline{g}(\psi,\alpha, \epsilon, \mu ) = \frac{1}{\tau_b} \int_{l_1}^{l_2} g^+(\psi, \alpha, l, \epsilon, \mu) \frac{dl}{| v_\| |}, $$
in the trapped region of phase space, $\lambda B_{\rm max} > 1$. Here $l_1$ and $l_2$ denote consecutive bounce points, $B(\psi,\alpha,l_1) = B(\psi, \alpha, l_2) = 1/\lambda$, and 
	$$ \tau_b = \int_{l_1}^{l_2}  \frac{dl}{| v_\| |} $$
the bounce time. For circulating particles, $\lambda > 1/B_{\rm max}$, the orbit average is defined as
	$$ \overline{g}(\psi, \epsilon, \mu, \sigma ) = \lim_{L \rightarrow \infty} \int_0^L g(\psi, \alpha, l, \epsilon, \mu) \frac{dl}{v_\| }  
	\bigg\slash \int_0^L \frac{dl}{v_\| }. $$
Since the integral extends along a field line many times around the torus, the result is independent of $\alpha$. It can also be written in terms of the flux-surface average $\lang \cdots \rang$ as [\cite{Helander-2014-a}]
	$$ \overline{g}(\psi, \epsilon, \mu, \sigma ) = \lang \frac{B g}{v_\|} \rang \bigg\slash \lang \frac{B}{v_\|} \rang. $$
Note that the ``orbit average'' thus defined is a time average over the trajectories that would result if the drift velocity were neglected in the guiding-centre equations of motion. The actual orbits are more complicated since particles can drift in and out of local trapping wells. 

The reason for introducing the orbit average as done here is that this operation annihilates the first term in Eq.\,(\ref{+}), which becomes 
	\bn \overline{{\bf v}_d \cdot \nabla f^+} = \overline{C^+(f)}. 
	\label{orbit-averaged +}
	\en
The flow velocity of each species within the flux surface is determined by the odd part of the distribution function, which is given by	the remainder of Eq.\,(\ref{+}),
	\bn f^-(\psi,\alpha,l,\epsilon,\mu,\sigma) = \int_{l_0}^l \left[C^+(f) - {\bf v}_d \cdot \nabla f^+ \right] \frac{dl'}{v_\|} + X(\psi,\alpha,\epsilon, \mu, \sigma), 
	\label{f-}
	\en
where $l_0$ is arbitrary for the time being, and $X$ is an integration constant that needs to be determined from the orbit average of Eq.\,(\ref{-}),
	\bn
	\overline{{\bf v}_d \cdot \nabla f^-} = \overline{C^-(f)}. 
		\label{orbit-averaged -}
	\en
Our remaining task is to simplify and solve these equations. Of course, since the beginnings of neoclassical transport theory, this has been accomplished before in various approximations. In particular, the $1/\nu$-regime can be treated in general stellarator geometry, but to our knowledge the bootstrap current has not been calculated before in the $\sqrt{\nu}$-regime. In order to do so, we must first carefully go through the conventional $1/\nu$-regime (until Section 5), recovering familiar results but carefully identifying the assumptions made on the way.

\subsection{Even part of the distribution function}
 
The ratio of the right-hand side to the left-hand side of Eq.\,(\ref{orbit-averaged +}) is of order $\nu_{\ast } / \rho_\ast $, where $\rho_\ast = \rho /L$ is the normalised gyroradius and $\nu_\ast = \nu L/v_T$ the collisionality. Here $L$ denotes the macroscopic system length (typically the major radius) and $v_T = (2T/m)^{1/2}$ the thermal velocity. If $\rho_\ast \ll \nu_\ast$, collisions dominate and the distribution function is forced to be close to a Maxwellian. However, at lower collisionality, $\nu_\ast \ll \rho_\ast$, the plasma will in general not be in local thermodynamic equilibrium, since some particle orbits drift out of the plasma, leaving behind a ``loss cone'' in velocity space [\cite{Helander-2014-a, Calvo-2016}]. We are aware of two limits in which the particles are well confined and confinement is restored. One limit is described by
\cite{Ho-1987}, who took the ${\bf E} \times {\bf B}$ drift, 
$${\bf v}_E = - \frac{1}{B} \nabla \Phi \times {\bf b} \simeq \frac{1}{B} \frac{d \Phi}{d \psi} {\bf b} \times \nabla \psi, $$
to be much larger than the $\nabla B$ and curvature drifts, 
$${\bf v}_M = \frac{v_\|^2}{\Omega} {\bf b} \times ({\bf b} \cdot \nabla {\bf b}) + \frac{\mu}{m \Omega} {\bf b} \times \nabla B,$$ 
with $\Omega$ the gyrofrequency, i.e.,
\begin{equation} \label{ho order}
{\bf v}_E \gg {\bf v}_M.
\end{equation}
This limit is consistent with large aspect ratio, but it is also valid in regions of large plasma gradients of a low-aspect-ratio stellarator. The other possible limit considered in the literature is that of a stellarator sufficiently optimised that it can be considered close to omnigeneous [\cite{Calvo-2016}]. In this limit, the orbit-average of the radial drift is small compared with its local value,
\begin{equation} \label{calvo order}
{\bf v}_E \sim {\bf v}_M , \quad ({\bf v}_E + {\bf v}_M) \cdot \nabla \psi \gg \overline{({\bf v}_E + {\bf v}_M) \cdot \nabla \psi}.
\end{equation}
In both these limits, the bounce-averaged orbits stay close to flux surfaces, orbit confinement is restored, and the distribution function becomes approximately Maxwellian and constant on flux surfaces. Moreover, quasineutrality leads to an electrostatic potential $\Phi(\psi)$ that is constant on flux surfaces to lowest order [\cite{Ho-1987, Calvo-2016}].

Thus, assuming that the plasma is either in the $1/\nu$-regime or follows the orderings (\ref{ho order}) or (\ref{calvo order}), we shall write the solution to Eq.\,(\ref{orbit-averaged +}) as
	\bn f^+ = F_0 + F_1, 
	\label{F+ split}
	\en
where $F_0(\epsilon, \psi)$ is a Maxwellian that is constant on flux surfaces, and $F_1 \ll F_0$ is constant along magnetic field lines, that is, independent of $l$ for trapped particles, and independent of $\alpha$ and $l$ for circulating particles. The piece $F_1$ is determined by 
	\bn \overline{{\bf v}_d \cdot \nabla \alpha}\, \frac{\partial F_1}{\partial \alpha} - \overline{C^+(F_1)} = - \overline{{\bf v}_d \cdot \nabla \psi}\, \frac{\partial F_0}{\partial \psi}. 
	\label{Eq for F+}
	\en
The solution to this equation depends on the collisionality and the radial electric field, which enters in ${\bf v}_d$. Since 
	$$ \overline{{\bf v}_d \cdot \nabla \psi}\, \frac{\partial F_0}{\partial \psi} = 0 $$
in the untrapped part of phase space, there is no ``drive'' there, and most simple collision operators, including Eq.\,(\ref{C}), will lead to $F_1 = 0$ for circulating particles. The full operator, however, includes a field-particle term that does not have this property [\cite{Helander-2002-c}], but we shall ignore this complication. The function $F_1$ in the circulating domain is small compared to $F_1$ in the trapped particle region if we use the assumptions in (\ref{ho order}) and (\ref{calvo order}) [\cite{Calvo-2016}]. 

Equation (\ref{Eq for F+}), with the approximation (\ref{C}) for the collision operator, is routinely solved by the GSRAKE code [\cite{Beidler-1995}], which uses the result to calculate the cross-field transport. This calculation is very rapid compared with codes that attempt to solve the drift kinetic equation without first taking the bounce average.

We consider two low-collisionality limits of Eq.\,(\ref{Eq for F+}). In the higher-collisionality limit (corresponding to the $1/\nu$-regime), $C(F_1) \gg \overline{{\bf v}_d \cdot \nabla \alpha}\; (\p F_1/\p \alpha)$. Thus, $\overline{{\bf v}_d \cdot \nabla \alpha}\; (\p F_1/\p \alpha) \ll \overline{{\bf v}_d \cdot \nabla \psi} \, (\p F_0/\p \psi)$, which in general implies 
	\bn {\bf v}_d \cdot \nabla F_1 \ll {\bf v}_d \cdot \nabla F_0.
	\label{high C}
	\en
In the lower-collisionality limit (corresponding to the $\sqrt{\nu}$-regime), 
$$C(F_1) \ll \overline{{\bf v}_d \cdot \nabla \alpha}\; (\p F_1/\p \alpha),$$ leading to 
	\bn C(F_1) \ll {\bf v}_d \cdot \nabla F_1 \sim {\bf v}_d \cdot \nabla F_0.
	\label{low C}
	\en
Sections 3-5 below deal with the $1/\nu$-regime and Section 6 with the $\sqrt{\nu}$-regime.

\subsection{Odd part of the distribution function}

The odd part of the distribution function is determined by Eqs.~(\ref{f-}) - (\ref{orbit-averaged -}) and must vanish at the bounce points. This condition alone determines the integration constant $X$ for trapped orbits; specifically, if we choose $l_0$ to be a bounce point then $X=0$. It is thus natural to let $l_0$ depend on $\lambda$ in the following way:
	\begin{eqnarray*} 
	B(l_0) = \left\{ \begin{array}{cc}
	1/\lambda, \qquad & \lambda > 1/B_{\rm max}, \\
	B_{\rm max}, \qquad & \lambda < 1/B_{\rm max}. 
	\end{array} \right.
	\end{eqnarray*}
In order to calculate $X$ for untrapped orbits, we must solve Eq.\,(\ref{orbit-averaged -}), which we write as
	\bn \lang \frac{B}{v_\|}  C^-(f) \rang = \lang \nabla \cdot \left( \frac{B f^-}{v_\|} {\bf v}_d \right)	 \rang
	= \frac{1}{V'(\psi)} \frac{\p}{\p \psi} \lang \frac{V'(\psi) B f^-}{v_\|} {\bf v}_d \cdot \nabla \psi \rang, 
	\label{Eq for X}
	\en
where $V(\psi)$ denotes the volume contained by the flux surface labelled by $\psi$, $V^\prime (\psi) = dV/d\psi$, and we have used ${\bf v}_d = (v_\| / \Omega) \nabla \times (v_\| {\bf b})$. On the right-hand side of this expression appears
	$$ \lang \frac{B f^-}{v_\|}  {\bf v}_d \cdot \nabla \psi \rang = \lang \frac{B {\bf v}_d \cdot \nabla \psi }{v_\|} \left[
	\int_{l_0}^l \left(C^+(f) - {\bf v}_d \cdot \nabla f^+ \right) \frac{dl'}{v_\|} + X \right] \rang, $$
where the last term vanishes because the constant $X$ is a flux function for untrapped orbits and [\cite{Helander-2014-a}]
	$$ \lang \frac{B {\bf v}_d \cdot \nabla \psi }{v_\|} \rang = 0. $$
Using Eq.\,(\ref{F+ split}) and the fact that $f^+ = F_0 + F_1$ does not depend on $\alpha$ or $l$ in the circulating-particle region, Eq.\,(\ref{Eq for X}) can be rewritten as
	\bn \lang \frac{B}{v_\|}  C^-(f) \rang = \frac{1}{V'(\psi)} \frac{\p}{\p \psi} V'(\psi) \lang \frac{B {\bf v}_d \cdot \nabla \psi}{v_\|} \int_{l_0}^l \left( C(F_1) - {\bf v}_d \cdot \nabla \psi \frac{\partial f^+}{\partial \psi} \right ) \frac{dl'}{v_\|} \rang. 
	\label{Eq for X v2}
	\en
If we write
	$$ h(l,\epsilon, \mu) = \int_{l_0}^l {\bf v}_d \cdot \nabla \psi \frac{dl'}{v_\|}, $$
then $\nabla_\| h = ({\bf v}_d \cdot \nabla \psi) / v_\|$ and
	$$ \lang \frac{B {\bf v}_d \cdot \nabla \psi }{v_\|} \frac{\partial f^+}{\partial \psi} \int_{l_0}^l {\bf v}_d \cdot \nabla \psi \frac{dl'}{v_\|} \rang = 
	\frac{\p f^+}{\p \psi} \lang \frac{B}{2} \nabla_\| h^2 \rang = 0, $$
since $\lang B \nabla_\| g \rang$ vanishes for any function $g$, see Eq.~(21) of \cite{Helander-2014-a}.
Equation (\ref{Eq for X v2}) thus reduces to 
	$$ \lang \frac{B}{v_\|}  C^-(f) \rang 
	= \frac{1}{V'} \frac{\p}{\p \psi} V' \lang \frac{B {\bf v}_d \cdot \nabla \psi}{v_\|} 
	\int_{l_0}^l C(F_1) \frac{dl'}{v_\|}  \rang, $$
where the right-hand side is small compared with the left-hand side, since 
	$$ \frac{1}{V'} \frac{\p}{\p \psi} V' \lang \frac{B {\bf v}_d \cdot \nabla \psi}{v_\|} 
	\int_{l_0}^l C(F_1)\frac{dl'}{v_\|} \rang \bigg\slash \lang \frac{B}{v_\|} C^-(f)\rang 
	\sim \frac{\rho_\ast F_1}{f^-} \sim \frac{F_1}{F_0} \ll 1, $$
and we have used $f^- \sim \rho_\ast F_0$, which is to be verified {\em a posteriori}. Furthermore, since we have seen that $F_1=0$ for untrapped particles (for the collision operator we employ), we can thus conclude that the equation determining the integration constant $X$ to lowest order is
	\bn \lang \frac{B}{v_\|}  C^-(f) \rang	= 0 
	\label{eq for X}
	\en
in all low-collisionality regimes. 

Once the integration constant has been obtained, it is straightforward to calculate the flow velocity along the magnetic field of the particle species in question,
	$$ V_{a\|}(l) = \frac{1}{n_a} \int v_\| f_a^- d^3v $$
	\bn
	= \frac{1}{n_a} \left( - B(l)\int_{l_0}^l \frac{dl'}{B(l')} 
	\int {\bf v}_d \cdot \nabla f_a^+ d^3v + \int v_\| X_a(\psi,\epsilon, \mu) \; d^3v \right) 
	= V_{a1} + V_{a2}. \label{V}
	\en
Note that the term with the collision operator does not contribute.
In the $1/\nu$ regime, according to Eq.\,(\ref{high C}), ${\bf v}_d \cdot \nabla f_a^+ \simeq {\bf v}_d \cdot \nabla F_0$. If we use this approximation and the expression [e.g., from \cite{Landreman-2012} or Eq.~(63) of \cite{Helander-2014-a}]
	$$ {\bf v}_{d} \cdot \nabla \psi = \frac{m_a v^2}{e_a} \xi \; ({\bf b} \times \nabla \psi) \cdot \nabla 
	\left( \frac{\xi}{B} \right), $$
where the gradient of $\xi = \sigma (1-\lambda B)^{1/2}$ is taken at constant $\lambda$, we obtain for the first term
	$$ \lang V_{a1} B \rang = - \lang \frac{m_a B^2}{n_a e_a} \int_0^\infty \frac{\p F_{a0}}{\p \psi} \; 2 \pi v^4 dv 
	\int_0^{1/B} d \lambda \int_{l_0}^l \left( {\bf b} \times \nabla \psi \right) \cdot \nabla \left( \frac{\xi }{B} \right) dl' \rang. $$ 
Evaluating the integrals, we find [\cite{Helander-2011-a}]
	\bn \lang V_{a1} B \rang = \frac{T_a A_{1a}}{e_a} \lang uB^2 \rang, 
	\label{V1}
	\en
where we have recalled Eq.\,(\ref{eq for u}) and fixed the integration constant therein by taking $u=0$ at the point on the flux surface where $B=B_{\rm max}$. (In the notation of \cite{Nakajima-1989} and \cite{Helander-2011-a}, $g_2 = - uB^2$.) In addition, we have defined the thermodynamic forces		
	$$ A_{1a} = \frac{d \ln p_a}{d\psi} + \frac{e_a}{T_a} \frac{d\Phi}{d\psi}, $$
	$$ A_{2a} = \frac{d \ln T_a}{d\psi}. $$
The approximation ${\bf v}_d \cdot \nabla f_a^+ \simeq {\bf v}_d \cdot \nabla F_{a0}$ is, however, not appropriate in all collisionality regimes and will be discussed further below, but first we shall solve Eq.\,(\ref{eq for X}) in various cases, still using this approximation, and calculate the corresponding velocities $V_{a2}$.

\section{Lorentz limit}

To solve Eq.\,(\ref{eq for X}), we must specify the collision operator. We first consider pure pitch-angle scattering, i.e., we retain only the first term on the right-hand side of Eq.\,(\ref{C}). This applies to electrons in a high-$Z$ plasma and is used in the widespread DKES code [\cite{Hirshman-1986,van-Rij-1989}]. 

In the coordinates $v$ and $\lambda$, the operator $\mathcal{L}$ becomes 
$$\mathcal{L}(f) = \frac{2\xi}{B} \frac{\p}{\p \lambda} \left ( \lambda \xi \frac{\p f}{\p \lambda} \right ).$$
Our equation (\ref{eq for X}) then becomes
	$$ \frac{\p}{\p \lambda} \lambda  \lang \xi \frac{\p f^-}{\p \lambda} \rang = 0, $$
and is readily integrated once, demanding regularity at $\lambda = 0$, 
	\bn \lang \xi \frac{\p f^-}{\p \lambda} \rang = 0. 
	\label{1st integral}
	\en
We substitute Eq.\,(\ref{f-}) and note that the term involving the collision operator vanishes in the untrapped part of phase space since $F_1 = 0$ there. Hence we obtain
	$$ \frac{\p X}{\p \lambda} = \frac{1}{\lang \xi \rang} \frac{\p F_0}{\p \psi} \lang \xi \frac{\p}{\p \lambda} \int_{l_{\rm max}}^l {\bf v}_d \cdot \nabla \psi \frac{dl'}{v_\|(l')} \rang, $$
where $B(l_{\rm max}) = B_{\rm max} $. We thus find
	\bn \frac{\p X_a}{\p \lambda} = - \frac{m_a v}{2 e_a} \frac{\p F_0}{\p \psi} \frac{\lang g_4 \rang}{\lang \xi \rang},
	\label{dX/dlambda}
	\en
where we have introduced the notation [\cite{Nakajima-1989}]
		$$ g_4(\lambda,l) = \xi \int_{l_{\rm max}}^l ({\bf b} \times \nabla \psi) \cdot \nabla \xi^{-1} dl', 
	\qquad (\lambda < 1/B_{\rm max}). $$
Knowing $X_a$, we can now calculate
		$$ V_{a2} = \frac{1}{n_a} \int v_\| X_a(\psi,\epsilon, \mu) \; d^3v = - \frac{1}{n_a} \int v_\| \lambda \frac{\p X_a}{\p \lambda}\; d^3 v,
		$$
which becomes
		$$  \lang V_{a2} B \rang = \frac{f_s T_a A_{1a}}{e_a}, $$
where we have written
		$$ f_s = \frac{3 \lang B^2 \rang}{4} \int_0^{1/B_{\rm max}} \frac{ \lang g_4 \rang \lambda d\lambda}{\lang \sqrt{1-\lambda B} \rang}. $$
The current along $\bf B$ is thus equal to  [\cite{Helander-2011-a}]
			\bn \lang J_{a\|} B \rang = n_a e_a \lang (V_{a1} + V_{a2}) B \rang =  p_a A_{1a} \left( f_s + \lang uB^2 \rang \right) 
			\label{Lorentz J}
			\en
in this approximation. Note that this current depends on the radial electric field in $A_{1a}$ but disappears, as it should, from the total current once a sum is taken over all particle species. However, there are perils of using a collision operator that does not conserve momentum. In a tokamak or quasisymmetric stellarator, the resulting transport depends on the radial electric field, but this dependence disappears if a momentum-conserving operator is used. In addition, there is spurious particle transport from self-collisions, which vanishes if momentum is conserved.
		
\section{Rotation velocity of a pure plasma}
	
A better approximation, particularly for the plasma ions ($a=i$), which are assumed to be singly charged, $e_i = e$, is obtained if the momentum-restoring term is retained in the collision operator (\ref{C}). For the bulk ions in a plasma without impurities, we thus take 
	\bn C_i^-(f_i) = \nu_D^{ii} \left( {\cal L} (f_i^-) + \frac{m_i v_\|}{T_i} w_i F_{i0} \right), 
	\label{Ci}
	\en
where 
	$$\nu_D^{ii} = \frac{3 \pi^{1/2}}{4 \tau_{ii}} \frac{\phi(x) - G(x)}{x^3}, $$
denotes the ion-ion pitch-angle scattering frequency, with 
  $$\tau_{ii} = 6\sqrt{2} \pi^{3/2} \epsilon_0^2 m_i^{1/2} T_i^{3/2}/e^4 n_i \ln \Lambda,$$ 
	$\ln \Lambda$ the Coulomb logarithm, $\epsilon_0$ the vacuum permittivity, $x=v/v_{Ti}$, $v_{Ti} = \sqrt{2 T_i / m_i}$ and 
	$$ \phi(x) = \frac{2}{\sqrt{\pi}} \int_0^x e^{-y^2} dy, $$
	$$ G(x) = \frac{\phi(x)-x\phi'(x)}{2x^2}. $$
Instead of Eq.\,(\ref{1st integral}) we now obtain
	$$ \lang \xi \frac{\p f_i^-}{\p \lambda} \rang = - \frac{m_i v}{2T_i} \lang w_i B \rang F_{i0}, $$
and instead of Eq.\,(\ref{dX/dlambda})
	\bn \frac{\p X_i}{\p \lambda} = - \frac{m_i v}{2 T_i \lang \xi \rang} \left( \lang w_i B \rang F_{i0} +  \frac{T_i \lang g_4 \rang}{e} \frac{\p F_{i0}}{\p \psi} \right).
	\label{dXi/dlambda}
	\en
The coefficient $w_i$ is determined by momentum conservation,
	$$ 0 = \int v_\| C_i^-(f_i) \; d^3v = n_i w_i \left\{ \nu_D^{ii} \right\} - \int \nu_D^{ii} v_\| f_i^- d^3v,  $$                                                      
where we have defined the velocity-space average [\cite{Helander-2002-c}]
		$$ \left\{ \nu_D^{ii} \right\} = \int \nu_D^{ii} \frac{m_i v^2}{3 n_i T_i} F_{i0} \; d^3v = \frac{\sqrt{2} - \ln \left( 1+ \sqrt{2} \right)}{\tau_{ii}}
		\simeq \frac{0.533}{\tau_{ii}}.$$
Upon substitution of Eq.\,(\ref{f-}) we find
	\bn \lang w_i B \rang = \frac{1}{n_i \left\{ \nu_D^{ii} \right\}} \lang B^2 \int_0^\infty \nu_D^{ii} \; 2\pi v^3 dv \int_0^{1/B} d\lambda
	\left( X_i - \int_{l_0}^l {\bf v}_d \cdot \nabla f_i^+ \frac{dl'}{|v_\| |} \right) \rang,
	\label{ui}
	\en
where we have noted that the part of the distribution function (\ref{f-}) containing $C^+(f)$ does not contribute to the integral:
	$$ \int \nu_D^{ii} v_\| \; d^3v \int_{l_0}^l C^+(f) \frac{dl'}{v_\|} = B(l) \int_{l_{\rm max}}^l \frac{dl'}{B(l')} \int \nu_D^{ii} C^+(f) \; d^3v = 0, $$
for any collision operator of the form (\ref{C}). 
	
The first of the two terms on the right of Eq.\,(\ref{ui}) can be evaluated using Eq.\,(\ref{dX/dlambda}) and is equal to 				
	$$  - \frac{1}{n_i \left\{ \nu_D^{ii} \right\}} \lang B^2 \int_0^\infty \nu_D^{ii} \; 2\pi v^3 dv \int_0^{1/B_{\rm max}} \lambda	\frac{\p X_i}{\p \lambda} d\lambda \rang
	= f_c \lang w_i B \rang + \frac{f_s T_i}{e} \left( A_{1i} - \eta A_{2i} \right),
	$$
with
	$$ f_c = \frac{3 \lang B^2 \rang}{4} \int_0^{1/B_{\rm max}} \frac{\lambda d\lambda}{\lang \sqrt{1-\lambda B} \rang}, $$
	$$ \eta = \frac{\left\{ \nu_D^{ii} \left(5/2 -  x^2\right) \right\} }{  \left\{ \nu_D^{ii} \right\}} = \frac{5}{2} - \frac{1}{2 - \sqrt{2} \ln (1 + \sqrt{2})} \simeq 1.17, $$
The second term on the right of Eq.\,(\ref{ui}) becomes 
	$$ - \frac{1}{n_i \left\{ \nu_D^{ii} \right\}} \lang B^2 \int_0^\infty \nu_D^{ii} \; 2\pi v^3 dv \int_0^{1/B} d\lambda 
	\int_{l_0}^l {\bf v}_d \cdot \nabla f_i^+ \frac{dl'}{|v_\| |} \rang =
	\frac{T_i \lang uB^2 \rang}{e} \left( A_{1i} - \eta A_{2i} \right), $$
where we have again approximated ${\bf v}_d \cdot \nabla f^+$ by ${\bf v}_d \cdot \nabla F_0$. As long as this approximation is allowed, Eq.\,(\ref{ui}) thus implies 
		\bn \lang w_i B \rang = \frac{T_i \left(f_s + \lang uB^2 \rang \right)}{e (1-f_c)} \left( A_{1i} - \eta A_{2i} \right),
		\label{final ui}
		\en
which together with Eq.\,(\ref{dXi/dlambda}) determines the integration constant $X_i$ and thus the odd part of the ion distribution function. 

We are now in a position to calculate the plasma flow velocity along the magnetic field, $\lang V_{i\|} B \rang = \lang (V_{i1} + V_{i2}) B \rang$. The first term was calculated in Eq.\,(\ref{V1}) and the second term is
	\bn \lang V_{i2} B \rang = - \frac{2 \pi}{n_i} \lang B^2 \int_0^\infty v^2 dv \int_0^{1/B_{\rm max}} \frac{\p X_i}{\p \lambda} \lambda d\lambda \rang 
	 = f_c \lang w_i B \rang + \frac{f_s T_i A_{1i}}{e}, 
	\label{V2}
	\en
where we have used Eq.\,(\ref{dX/dlambda}). 
We thus obtain
		\bn \lang V_{i\|} B \rang = \frac{T_i}{e} \frac{f_s + \lang u B^2 \rang}{1-f_c} \left( A_{1i} - \eta f_c A_{2i} \right). 
		\label{Vi}
		\en
Apart from the geometric factor, this result is similar to that in tokamaks and was obtained by \cite{Landreman-2012} for the case of a perfectly omnigenous stellarator. From our analysis, it is clear that the same result is also valid in the $1/\nu$-regime of any non-optimised stellarator. A similar conclusion follows from the moment method of Hirshman and Sigmar extended to stellarator geometry by \cite{Shaing-1983}, in whose paper this expression can also be found. A refined expression is obtained in the next section using a more accurate collision operator. We note, however, that it is only valid if the ${\bf v}_d \cdot \nabla f_i^+$ can indeed be approximated by  ${\bf v}_d \cdot \nabla F_{i0}$ in Eq.\,(\ref{f-}). As we shall see below, the correction that arises in the $\sqrt{\nu}$-regime can be significant. 

In an axisymmetric magnetic field, ${\bf B} = F(\psi) \nabla \varphi + \iota \nabla \varphi \times \nabla \psi$, 
		$$ f_s = \frac{F}{\iota} \left( f_c - \frac{\lang B^2 \rang}{B_{\rm max}^2} \right), $$
and 
	$$ \lang u B^2 \rang = \frac{F}{\iota} \left( \frac{\lang B^2 \rang}{B_{\rm max}^2} - 1 \right), $$
so that $f_s + \lang u B^2 \rang = - (1-f_c) F / \iota$. The ion flow velocity (\ref{Vi}) thus reduces to the usual result [\cite{Helander-2002-c}]
	$$ \lang V_{i\|} B \rang = - \frac{F}{\iota}\frac{T_i}{e} \left( A_{1i} - \eta f_c A_{2i} \right). $$
	
\section{Bootstrap current}

To calculate the bootstrap current, we also need to determine the distribution function for the electrons. In order to obtain a result that matches that for a standard large-aspect-ratio tokamak [\cite{Rosenbluth-1972, Helander-2002-c}], we use a model linearised electron collision operator $C_e(f_e)$ that is more accurate than the operator in Eq.\,(\ref{Ci}). \cite{Rosenbluth-1972} showed that the operator (\ref{C}) is sufficient in the limit $f_t = 1 - f_c \rightarrow 0$, and it later emerged that improved accuracy can be achieved at finite $f_t$ by letting the momentum-restoring coefficient $w$ depend on energy. We therefore employ such an operator as the most accurate one allowing for an explicit analytical solution of the drift kinetic equation, but remind the reader that its accuracy is not perfect (\cite{Parker-2012}).

The electron collision operator must include both electron-electron and electron-ion collisions,
	\bn C_e^-(f_e) = C_{ee}(f_e^-) + C_{ei} (f_e^-).
	\label{Ce}
	\en
To lowest order in $\sqrt{m_e/m_i} \ll 1$, the electron-ion collision operator is
	\bn C_{ei}(f_e^-) = \nu_D^{ei} \mathcal{L} ( f_e^-) + \frac{\nu_D^{ei} m_e v_\| V_{i\|}}{T_e} F_{e0},
	\label{Cei}
	\en
where $\nu_D^{ei} = (3  \sqrt{\pi} /4 \tau_{ei} x^3)$, $x = v/v_{Te}$, $v_{Te} = \sqrt{2T_e/m_e}$ and, for singly charged ions, $\tau_{ei} = \tau_{ee}= 6\sqrt{2} \pi^{3/2} \epsilon_0^2 m_e^{1/2} T_e^{3/2}/e^4 n_e \ln \Lambda$. For $C_{ee}$, we use a model collision operator [\cite{Taguchi-1988,Helander-2002-c}] that has the exact pitch-angle scattering piece of the Landau collision operator, conserves momentum, is self-adjoint, has $f_e^- = (m_e v_\| V_{e\|}/T_e) F_{e0}$ as the only solution to $C_{ee} (f_e^-) = 0$, and in addition, we request that it satisfies 
	\bn
	\int L_p^{(3/2)} (x^2) v_\| C_{ee} \left ( L_q^{(3/2)} (x^2) v_\| F_{e0} \right )\, d^3 v = - \frac{K_{pq} n_e T_e}{m_e \tau_{ee}}
	\label{Sonine condition}
	\en
for $p,q = 0, 1, \ldots, S$. Here $L_p^{(3/2)}(y)$ are Sonine polynomials, and $K_{pq}$ are coefficients that result from calculating $\int L_p^{(3/2)} (x^2) v_\| C_{ee} \left ( L_q^{(3/2)} (x^2) v_\| F_{e0} \right )\, d^3 v$ using the linearised Landau electron-electron collision operator. The first few Sonine polynomials are
	$$L_0^{(3/2)}(y) = 1, \quad L_1^{(3/2)}(y) = \frac{5}{2} - y, \quad L_2^{(3/2)}(y) = \frac{35}{8} - \frac{7 y}{2} + \frac{y^2}{2},$$
and the corresponding coefficients $K_{pq}$ are
	$$K_{p0} = K_{0q} = 0, \quad K_{11} = \sqrt{2}, \quad K_{12} = K_{21} = \frac{3\sqrt{2}}{4}, \quad K_{22} = \frac{45\sqrt{2}}{16}.$$
We derive an operator that satisfies these conditions in Appendix~\ref{app:Cee}. For our calculation, it is sufficient to satisfy condition (\ref{Sonine condition}) for $p, q = 0, 1, 2$. Thus, we use the model collision operator
	\bn C_{ee} (f_e^-) = \nu_D^{ee} \mathcal{L} ( f_e^- ) + \frac{\nu_D^{ee} m_e v_\|}{T_e} w_e F_{e0},
	\label{Cee}
	\en
where $\nu_D^{ee} = (3 \sqrt{\pi} /4 \tau_{ee}) [ \phi(x) - G(x) ]/x^3$, and $w_e$ is not a constant, but a function of $v$ given by
	$$w_e = a_0 + a_1 L_1^{(3/2)}(x^2) + a_2 L_2^{(3/2)}(x^2).$$
The coefficients $a_0$, $a_1$ and $a_2$ are determined by the linear set of equations
	\bn \sum_{q = 0}^2 N_{pq} a_q= \frac{\tau_{ee}}{n_e} \int \nu_D^{ee} L_p^{(3/2)}(x^2) v_\| f_e^-\, d^3 v,
	\label{Eq for a}
	\en
where the symmetric matrix $N_{pq}$ is given by
	$$\left ( \begin{array}{c c c}
N_{00} & N_{01} & N_{02} \\
N_{10} & N_{11} & N_{12} \\
N_{20} & N_{21} & N_{22}
\end{array} \right ) = \left ( \begin{array}{c c c}
0.533 & 0.625 & 0.652 \\
0.625 & 0.0240 & -0.288 \\
0.652 & -0.288 & -0.947
\end{array} \right ).$$
In Appendix~\ref{app:Cee} we show how to calculate this matrix.

Using the collision operators in Eqs. (\ref{Ce}), (\ref{Cei}) and (\ref{Cee}), we find that
	\bn \frac{\partial X_e}{\partial \lambda} = - \frac{m_e v}{2T_e \langle \xi \rangle} \Bigg [ \sum_{p = 0}^2 \frac{\nu_D^{ee} L_p^{(3/2)}(x^2) \langle a_p B \rangle}{\nu_D^{ee} + \nu_D^{ei}} F_{e0}
	+ \frac{\nu_D^{ei} \langle V_{i\|} B \rangle}{\nu_D^{ee} + \nu_D^{ei}} F_{e0}
	- \frac{T_e \langle g_4 \rangle}{e} \frac{\partial F_{e0}}{\partial \psi} \Bigg ],
	\label{dXe/dlambda}
	\en
where the coefficients $\langle a_p B \rangle$ are determined from the flux surface average of Eqs. (\ref{Eq for a}) multiplied by $B$:
	\bea \sum_{q = 0}^2 \left ( N_{pq} - f_c \tau_{ee} \left \{ \frac{(\nu_D^{ee})^2 L_p^{(3/2)} L_q^{(3/2)}}{\nu_D^{ee} + \nu_D^{ei}} \right \} \right ) \langle a_q B \rangle =  \tau_{ee} \left \{ \frac{\nu_D^{ee} \nu_D^{ei} L_p^{(3/2)}}{\nu_D^{ee} + \nu_D^{ei}} \right \} f_c \langle V_{i\|} B \rangle \\ - \frac{T_e \tau_{ee} }{e} (f_s + \langle u B^2 \rangle) \Bigg [  \{ \nu_D^{ee} L_p^{(3/2)} \} A_{1e} - \left \{ \nu_D^{ee} L_1^{(3/2)} L_p^{(3/2)} \right \} A_{2e} \Bigg ].
	\eea
Using
\begin{eqnarray} \label{eq:coefficients1}
\left \{ \frac{(\nu_D^{ee})^2}{\nu_D^{ee} + \nu_D^{ei}} \right \} = \frac{0.198}{\tau_{ee}}, \quad \left \{ \frac{(\nu_D^{ee})^2 L_1^{(3/2)}}{\nu_D^{ee} + \nu_D^{ei}} \right \} = \frac{0.197}{\tau_{ee}}, \quad \left \{ \frac{(\nu_D^{ee})^2 L_2^{(3/2)}}{\nu_D^{ee} + \nu_D^{ei}} \right \} = \frac{0.179}{\tau_{ee}}, \nonumber\\
\left \{ \frac{(\nu_D^{ee})^2 \left ( L_1^{(3/2)} \right )^2}{\nu_D^{ee} + \nu_D^{ei}} \right \} = \frac{0.458}{\tau_{ee}}, \quad \left \{ \frac{(\nu_D^{ee})^2 L_1^{(3/2)} L_2^{(3/2)}}{\nu_D^{ee} + \nu_D^{ei}} \right \} = \frac{0.457}{\tau_{ee}}, \nonumber\\
\left \{ \frac{(\nu_D^{ee})^2 \left ( L_2^{(3/2)} \right )^2}{\nu_D^{ee} + \nu_D^{ei}} \right \} = \frac{0.754}{\tau_{ee}},  \nonumber\\ 
\left \{ \frac{\nu_D^{ee} \nu_D^{ei}}{\nu_D^{ee} + \nu_D^{ei}} \right \} = \frac{0.335}{\tau_{ee}}, \quad \left \{ \frac{\nu_D^{ee} \nu_D^{ei} L_1^{(3/2)}}{\nu_D^{ee} + \nu_D^{ei}} \right \} = \frac{0.428}{\tau_{ee}}, \quad \left \{ \frac{\nu_D^{ee} \nu_D^{ei} L_2^{(3/2)}}{\nu_D^{ee} + \nu_D^{ei}} \right \} = \frac{0.473}{\tau_{ee}},
\end{eqnarray}
we find the equations
	\begin{eqnarray}
	\left ( \begin{array}{c c c}
0.335 + 0.198f_t & 0.428 + 0.197f_t & 0.473 + 0.179f_t \\ 
0.428 + 0.197f_t  & -0.434 + 0.458f_t & -0.746 + 0.457f_t  \\
0.473 + 0.179f_t& -0.746 + 0.457f_t & -1.701 + 0.754f_t
\end{array} \right ) \left ( \begin{array}{c} 
\langle a_0 B \rangle \\ \langle a_1 B \rangle \\ \langle a_2 B \rangle
\end{array} \right ) \nonumber\\
= \frac{T_e}{e} (f_s + \langle u B^2 \rangle) \left [ \left ( \begin{array}{c}
- 0.533\\ -0.625\\-0.652
\end{array} \right ) A_{1e} + \left ( \begin{array}{c}
0.625\\1.386 \\1.563 
\end{array} \right )A_{2e} \right ] 
+ \left ( \begin{array}{c}
0.335\\ 0.428\\ 0.473
\end{array} \right ) f_c \langle V_{i\|} B\rangle.
	\label{linear}
	\end{eqnarray}  
Solving Eqs. (\ref{linear}), we find
	\begin{eqnarray}
	\left ( \begin{array}{c}
\langle a_0 B \rangle \\  \langle a_1 B \rangle \\  \langle a_2 B \rangle
\end{array} \right ) = \left ( \begin{array}{c}
-1.54 + 1.34 f_t - 0.22 f_t^2 \\
0.01 (1 - f_t^2) \\
-0.05 + 0.07 f_t - 0.02 f_t^2
\end{array} \right ) \frac{T_e}{eD} (f_s + \langle u B^2 \rangle) A_{1e} \nonumber\\ + \left ( \begin{array}{c}
2.72 - 1.63 f_t + 0.03 f_t^2 \\
-0.95 - 0.20 f_t + 0.21 f_t^2 \\
0.25 + 0.08 f_t + 0.06 f_t^2
\end{array} \right )\frac{T_e}{eD} (f_s + \langle u B^2 \rangle)  A_{2e} \nonumber\\ + \left ( \begin{array}{c}
1 - 0.86 f_t + 0.12 f_t^2 \\
0.01 f_t (1 + f_t) \\
f_t (- 0.05 + 0.02 f_t)
\end{array} \right ) \frac{f_c}{D} \langle V_{i\|} B \rangle, 
	\label{avea}
	\end{eqnarray}
where
	$$D = 1 - 0.32 f_t - 0.36 f_t^2 + 0.09 f_t^3.$$
	
With these results, we can red now obtain the  electron current. Using Eqs. (\ref{V}) and (\ref{dXe/dlambda}), we find
 	\bea \langle J_\| B \rangle = n_e T_e (f_s + \langle u B^2 \rangle) A_{1e} -e n_e \sum_{p=0}^2 \left \{ \frac{\nu_D^{ee} L_p^{(3/2)}}{\nu_D^{ee} + \nu_D^{ei}} \right \} f_c \langle a_p B \rangle \nonumber\\ + e n_e \left \{ \frac{\nu_D^{ee} + f_t \nu_D^{ei}}{\nu_D^{ee} + \nu_D^{ei}} \right \}\langle V_{i\|} B \rangle,
	\eea
where
	\bea \left \{ \frac{\nu_D^{ee}}{\nu_D^{ee} + \nu_D^{ei}} \right \} = 0.428, \quad \left \{ \frac{\nu_D^{ee} L_1^{(3/2)}}{\nu_D^{ee} + \nu_D^{ei}} \right \} = - 0.0563, \\
	\quad \left \{ \frac{\nu_D^{ee} L_2^{(3/2)}}{\nu_D^{ee} + \nu_D^{ei}} \right \} = -0.0430,
\left \{ \frac{\nu_D^{ei}}{\nu_D^{ee} + \nu_D^{ei}} \right \} = 0.572.
	\eea
Using Eq.\,(\ref{avea}), we thus obtain
	\begin{eqnarray}
	\langle J_\| B \rangle = \frac{1.66 - 1.55 f_t + 0.30 f_t^2}{D} \left [ n_e T_e (f_s + \langle u B^2 \rangle) A_{1e} + f_t e n_e \langle V_{i\|} B \rangle \right ] \nonumber\\ - \frac{(1.21 - 0.69 f_t) f_c}{D} n_e T_e (f_s + \langle u B^2 \rangle) A_{2e},
	\label{bootstrap}
	\end{eqnarray}

This result has been obtained with a more sophisticated collision operator	than the ion flow (\ref{Vi}). To refine the latter, allowing for an energy-dependent momentum-restoring coefficient,
	$$w_i = b_0 + b_1 L_1^{(3/2)}(x^2) + b_2 L_2^{(3/2)}(x^2),$$
we employ an operator similar to Eq.~(\ref{Cee}) for ion-ion collisions, which results in algebra very similar to (but simpler than) that for the electron current. Instead of Eq.~(\ref{dXe/dlambda}) one obtains
	$$ \frac{\partial X_i}{\partial \lambda} = - \frac{m_i v}{2T_i \langle \xi \rangle} \left[ \sum_{p = 0}^2 L_p^{(3/2)}(x^2) \langle b_p B \rangle F_{i0}
	+ \frac{T_i \langle g_4 \rangle}{e} \frac{\partial F_{i0}}{\partial \psi} \right], $$
and in place of the next equation
		\bea \sum_{q = 0}^2  \left( N_{pq} - f_c \tau_{ii}  \left\{ \nu_D^{ii} L_p^{(3/2)} L_q^{(3/2)} \right\}\right) \langle b_q B \rangle \\
	 	=  \frac{T_i \tau_{ii} }{e} (f_s + \langle u B^2 \rangle)
		\left[  \{ \nu_D^{ii} L_p^{(3/2)} \} A_{1i} - \left \{ \nu_D^{ii} L_1^{(3/2)} L_p^{(3/2)} \right \} A_{2i} \right].
	\eea
Using the velocity-space averages
	\begin{eqnarray*}
	\left\{ \nu_D^{ii} \right\} = \frac{0.533}{\tau_{ii}}, \quad 
	\left\{ \nu_D^{ii} L_1^{(3/2)} \right\} = \frac{0.625}{\tau_{ii}}, \quad 
	\left\{ \nu_D^{ii} L_2^{(3/2)} \right\} = \frac{0.652}{\tau_{ii}}, \\
	\left\{ \nu_D^{ii} \left( L_1^{(3/2)} \right)^2 \right\} = \frac{1.386}{\tau_{ii}}, \quad 
	\left\{ \nu_D^{ii} L_1^{(3/2)}L_2^{(3/2)} \right\} = \frac{1.563}{\tau_{ii}} \\
	\left\{ \nu_D^{ii} \left( L_2^{(3/2)} \right)^2 \right\} = \frac{2.484}{\tau_{ii}}, 
	\end{eqnarray*}
one finds the following system of equations
	\begin{eqnarray*}
	\left( \begin{array}{c c c}
0.533 f_t & 0.625 f_t & 0.652 f_t \\ 
0.625 f_t  & {-1.362 + 1.386 f_t} & {-1.852 + 1.563 f_t}  \\
0.652 f_t & {-1.852 +1.563 f_t} & {-3.431 + 2.484 f_t}
\end{array} \right) 
\left( \begin{array}{c} 
\langle b_0 B \rangle \\ \langle b_1 B \rangle \\ \langle b_2 B \rangle
\end{array} \right) 
\\
= \frac{T_i}{e} (f_s + \langle u B^2 \rangle) \left[ \left ( \begin{array}{c}
0.533\\ 0.625\\0.652
\end{array} \right ) A_{1i} - \left ( \begin{array}{c}
0.625\\1.386 \\1.563 
\end{array} \right )A_{2i} \right] 
\end{eqnarray*}
and the ion flow
 $$ \lang BV_{i\parallel} \rang = \frac{T_i}{e}  \frac{\lang uB^2 \rang + f_s}{f_t} \left( A_{1i} 
- {\frac{(1.17 - 0.89f_t)f_c}{1 - 1.27 f_t + 0.37 f_t^2}}A_{2i}\right).
$$

The final expression for the bootstrap current results from substituting this result in Eq.~(\ref{bootstrap}), giving
	$$
	\langle J_\| B \rangle = n (f_s + \lang uB^2 \rang) \frac{1.66 - 1.55f_t + 0.30 f_t^2}{D}  $$
 \bn \times \left[T_e A_{1e} - {\frac{(0.73 - 0.42f_t)f_c}{1 -0.94f_t + 0.18f_t^2}}T_e A_{2e} + T_i A_{1i} 
- {\frac{(1.17 - 0.89f_t)f_c}{1 - 1.27 f_t + 0.37 f_t^2}}T_iA_{2i}\right].
	\label{bootstrap2}
	\en
Note that the radial electric field terms present in the thermodynamic forces $A_{1i}$ and $A_{1e}$ cancel, so that the bootstrap current is independent of the electric field in the $1/\nu$-regime. To lowest order in $f_t$, this expression coincides with that of \cite{Shaing-1989}, who included two instead of three Sonine polynomials in the expansion.\footnote{Otherwise there would presumably have been exact agreement. The flux-surface average of the geometry function in Eq.~(75b) of \cite{Shaing-1983} which is needed to compare the results can be written in our notation as $\langle \tilde{G}_b \rangle = G  - \langle uB^2\rangle$, where $ G  = -[f_c\langle uB^2\rangle + f_s]/(1-f_c)$.} It also agrees with the result of \cite{Landreman-2012}, which it generalises to non-omnigenous stellarators in the $1/\nu$-regime.

In a large-aspect-ratio tokamak, the result (\ref{bootstrap2}) reduces to 
	$$ \langle J_\| B \rangle = - 1.66 f_t  \iota^{-1} F(\psi) n \left( T_e A_{1e} + T_i A_{1i} - 0.73 T_e A_{2e} - 1.17 T_i A_{2i} \right), $$
which closely approximates the standard expression [\cite{Rosenbluth-1972, Helander-2002-c}].

\section{Corrections due to tangential drifts}

We are now in a position to address our main objective, to calculate the flow velocity of any plasma species that is in the $\sqrt{\nu}$-regime of collisionality. This is practically always the case for the bulk ions in sufficiently hot stellarator plasmas. 

In our analysis above, we have made three approximations. First, the normalised gyroradius $\rho_\ast$ and collisionality $\nu_\ast$ have been assumed to be small. Second, it has been assumed that the distribution function is approximately Maxwellian, $f^+ = F_0 + F_1$ with $F_1 \ll F_0$. In an omnigenous magnetic field, this would follow from $\rho_\ast \ll 1$, and in a general stellarator it holds if $\rho_\ast \ll \nu_\ast$, but if $\nu_\ast$ is very small it requires that the magnetic field is sufficiently well optimised or that a radial electric field arises to confine the orbits [see the orderings in Eqs. (\ref{ho order}) and (\ref{calvo order})]. Our third approximation is that we have, on a couple of occasions, replaced ${\bf v}_d \cdot \nabla f^+$ by ${\bf v}_d \cdot \nabla F_0$ in Eq.\,(\ref{f-}). This approximation is justified in the $1/\nu$-regime [see Eq.\,(\ref{high C})] but not in the $\sqrt{\nu}$-regime. For very low collisionality, the correct ordering is given by Eq.\,(\ref{low C}), according to which Eq.\,(\ref{Eq for F+}) becomes 
	$$\overline{{\bf v}_d \cdot \nabla \alpha}\, \frac{\partial F_1}{\partial \alpha} 
	+ \overline{{\bf v}_d \cdot \nabla \psi}\, \frac{\partial F_0}{\partial \psi} = 0. $$
Note that collisions can now be neglected since their frequency is lower than that of the poloidal drift.\footnote{The collision operator is however important in regions where $\overline{{\bf v}_d \cdot \nabla \alpha} = 0$ and in thin boundary layers located at the boundaries between trapped and passing particles and at boundaries between particles trapped in different magnetic wells. The width of this boundary layer is of order $\Delta \xi \sim (\nu/|\Phi'(\psi)|)^{1/2}$ \cite{Ho-1987,Helander-2014-a,Calvo-2016}.} Using this equation for $F_1$ and employing either ordering (\ref{ho order}) or ordering (\ref{calvo order}), we find
	\bn {\bf v}_d \cdot \nabla f^+ = ( {\bf v}_d\cdot \nabla\psi - \overline{{\bf v}_d \cdot \nabla\psi} ) \frac{\partial F_0}{\partial \psi}.
	\label{vd low C}
	\en
In the ordering (\ref{calvo order}), the orbit average $\overline{{\bf v}_d \cdot \nabla\psi}$ is negligible compared to ${\bf v}_d\cdot \nabla\psi$, but in the ordering (\ref{ho order}) these two terms are comparable and ${\bf v}_d \cdot \nabla \alpha = \Phi'(\psi)$ is equal to its bounce average. To be as general as possible, we keep $\overline{{\bf v}_d \cdot \nabla\psi}$. We thus expect that the expressions for the plasma rotation velocity and bootstrap current velocity derived above may not be accurate at collisionalities below that of the $1/\nu$-regime. 
	
Equations (\ref{f-}), (\ref{low C}) and (\ref{vd low C}) then imply
	$$ f^- = -\int_{l_0}^l ({\bf v}_d \cdot \nabla \psi - \overline{{\bf v}_d \cdot \nabla \psi} ) \frac{\p F_0}{\p \psi} \frac{dl'}{v_\|} + X, $$
where $\overline{{\bf v}_d \cdot \nabla \psi}$ vanishes for circulating particles and $X$ for trapped ones. The odd part of the distribution function has thus acquired a new term, which one expects will tend to partly cancel the term containing ${\bf v}_d \cdot \nabla F_0$. This term vanishes for circulating particles and is small for trapped particles in a nearly omnigenous stellarator, but in general it can make a difference. In particular, if magnetic geometry is such that the radial drift velocity does not vary much over a bounce orbit, ${\bf v}_d \cdot \nabla \psi \simeq \overline{{\bf v}_d \cdot \nabla \psi}$, then $f^-$ (and the plasma flow velocity associated with it) will be considerably smaller than before. This is, for instance, the case in the limit considered in 
[\cite{Ho-1987}] of a stellarator with a large number of local magnetic wells.

Since $\overline{{\bf v}_d \cdot \nabla \psi}$ vanishes for circulating particles, this new term does not affect the analysis leading up to (\ref{eq for X}), but it does modify Eq.\,(\ref{V1}), which is replaced by
	$$ \lang V_{a1} B \rang = - \frac{1}{n_a} \lang {B^2} \int_{l_0}^l \frac{dl'}{B(l')} \int \left({\bf v}_d \cdot \nabla \psi 
	- \overline{{\bf v}_d \cdot \nabla \psi} \right)  \frac{\p F_0}{\p \psi} d^3v \rang = \frac{T_a A_{1a}}{e_a} (\lang u B^2 \rang + \beta), $$
with
		\bn \beta =  \lang \frac{3B^2}{2} \int_{1/B_{\rm max}}^{1/B} 
		\overline{ \xi ({\bf b} \times \nabla \psi) \cdot \nabla \left( \frac{\xi}{B} \right)} \; d\lambda 
		\int_{l_0}^l \frac{dl'}{\sqrt{1-\lambda B(l')}} \rang. 
		\label{beta}
		\en

Whether the new term affects the integration constant $X$ and thus the flow velocity $ V_{a2} $ in Eq.\,(\ref{V}) depends on the collision operator. In the case of pure Lorentz scattering, it does not and the current thus becomes equal to  
	$$ \lang B J_{a\|} \rang = \left( f_s + \lang uB^2 \rang + \beta \right) p_a A_{1a}.$$
instead of Eq.\,(\ref{Lorentz J}). 

If the momentum-restoring term is kept in the collision operator (\ref{C}), the integration constant $X$ is still given by Eq.\,(\ref{dXi/dlambda}) but is nevertheless affected by the radial electric field through $w_i$. In Eq.\,(\ref{ui}), we need to retain ${\bf v}_d \cdot \nabla F_1$, so that
	\begin{eqnarray}
	\lang w_i B \rang = - \frac{1}{n_i \left\{ \nu_D^{ii} \right\}} \Bigg \langle B^2 \int_0^\infty \nu_D^{ii} \; 2\pi v^3 dv \int_0^{1/B} d\lambda
	\Bigg ( \lambda \frac{\p X_i}{\p \lambda} \nonumber \\+ \int_{l_0}^l \left({\bf v}_d \cdot \nabla \psi - \overline{{\bf v}_d \cdot \nabla \psi} \right) 
	\frac{\p F_{i0}}{\p \psi} \frac{dl'}{|v_\| |} \Bigg ) \Bigg \rangle.
	\label{ui2}
	\end{eqnarray}
Evaluating the integrals on the right using (\ref{dXi/dlambda}) and (\ref{beta}) gives
	$$ \lang w_i B \rang = \frac{T_i \left(f_s + \lang uB^2 \rang  + \beta \right)}{e (1-f_c)} \left( A_{1i} - \eta A_{2i} \right). $$
The piece of the flow $\lang V_{i2} B \rang$ is still given by Eq.\,(\ref{V2}), and the ion flow velocity becomes
	$$ \lang V_{i\|} B \rang = \frac{T_i \left(f_s + \lang uB^2 \rang  + \beta \right)}{e (1-f_c)} \left( A_{1i} - \eta f_c A_{2i} \right), $$
instead of Eq.\,(\ref{Vi}). If the stellarator is well optimised in the sense defined by the ordering (\ref{calvo order}), $\overline{{\bf v}_d \cdot \nabla \psi}$ is small and, as a consequence, the correction $\beta$ will be unimportant. However, in other situations, $\beta$ will be as large as $f_s + \lang u B^2 \rang$, even if the fraction of trapped particles is small. Although the new term containing $\overline{{\bf v}_d \cdot \nabla \psi}$ only affects the distribution function in the trapped part of phase space, the response of the circulating particles is modified through the integration constant $X_i$. Plasma rotation is thus expected to be different in the $1/\nu$- and $\sqrt{\nu}$-regimes. 

Given that the bootstrap current in Eq.\,(\ref{bootstrap}) is proportional to $\lang V_{i\|} B \rang$, there will be a difference in the bootstrap current as well. In particular, the cancellation of the radial electric field does not occur in the $\sqrt{\nu}$ regime, leading to a bootstrap current that depends on $d \Phi/d\psi$. In plasmas where the electrons are so much hotter than the ions that the radial electric field is determined by the ``electron root'' of the ambipolarity condition equation, not only the ions, but also the electrons are in the $\sqrt{\nu}$-regime, and there will be an additional effect of the radial electric field on the bootstrap current.

\section{Finite collisionality}

In the limit of very small collisionality, the bootstrap current is independent of $\nu_\ast$, but numerical calculations reveal that finite-$\nu_\ast$ corrections are important in practice [\cite{Beidler-2011}]. \cite{Hinton-1973} showed that the first correction to the bootstrap current in a tokamak is proportional to $\nu_\ast^{1/2}$ and therefore cannot be calculated by simply continuing the expansion of the drift kinetic equation to higher order in $\nu_\ast$. Instead, a boundary layer forms in phase space between the regions of trapped and circulating orbits, and the correction to the bootstrap current scales as the width of this layer. Essentially the same calculation holds in an omnigenous stellarator [\cite{Helander-2011-a}], partly thanks to the fact that all maxima of the field strength on the same magnetic surface are similar [\cite{Cary-1997,Parra-2015}]. However, the problem is much more complicated in a stellarator of arbitrary geometry, where along a magnetic field line there are local maxima that are arbitrarily close but not equal to the global maximum, $B_{\rm max}$, making the structure of the boundary layer different from that in an axisymmetric field. 

We cannot perform a rigorous calculation like that of Hinton and Rosenbluth for a general stellarator, but we can estimate the width of the boundary layer by considering the statistical properties of the local magnetic-field-strength maxima. In a tokamak, where all such maxima are equal and equidistant, the estimate is obtained by noting that the effective collision frequency for scattering of $\xi = v_\|/v$ by an amount $\Delta \xi$ is $\nu_{\rm eff} \sim \nu / \Delta \xi^2$. 
The mean free path for such scattering is thus 
	\bn L_{\rm eff} = R \Delta \xi^2 / \nu_\ast, 
	\label{mfp}
	\en
where
	$$ \nu_\ast = \frac{R \nu}{v_T}. $$
The distance between two consecutive bounce points close to the trapped-passing boundary is $2 \pi q R$ in a tokamak of major radius $R$ and rotational transform $\iota = 1/q$, and is therefore equal to the mean free path for scattering by an amount
	$$ \Delta \xi \sim \sqrt{q \nu_\ast}. $$
This is the width of the collisional boundary layer around the trapped-passing boundary: particles within this region do not ``know'' whether they are trapped since collisional scattering across the boundary occurs on a time scale shorter than that of the bounce motion.

In order to investigate the stellarator case, we need to recognise that the local maxima of $B$ along a field line are in general all unequal. To calculate their distribution, we first choose magnetic coordinates on the flux surface we consider such that the maximum field strength is attained in the point $(\theta, \varphi) = (0,0)$\footnote{Here $\theta$ and $\varphi$ denote field-aligned magnetic coordinates whose origin is chosen to coincide with the magnetic-field maximum.}, which for simplicity is assumed to be the only point where $B=B_{\rm max}$.\footnote{In some stellarators, the maximum field strength is attained in an even number of points, which are stellarator-symmetrically located above and below the midplane. This should not affect the basic scaling derived here.} In its vicinity,
	$$ B(\theta, \varphi) \simeq B_{\rm max} + \frac{1}{2} \left( B_{\theta \theta} \theta^2 + 2 B_{\theta \varphi} \theta \varphi + B_{\varphi \varphi} \varphi^2 \right), $$
where derivatives are denoted by subscripts. Let us now follow $B(l)$ along a field line emanating from the point $(\theta, \varphi) = (0,0)$. If the rotational transform $\iota=n/m$ is a rational number, $B$ will reach $B_{\rm max}$ after $m$ toroidal turns, and if $\iota$ is irrational it will never do so. In any case, it will come close to $B=B_{\rm max}$ only when the field line returns to the neighbourhood of its starting point, i.e., when $\theta$ and $\varphi$ are close to zero (or multiples of $2 \pi$, which we may subtract). After $m$ toroidal turns around the torus, the value of the poloidal angle is
	$$ \theta_m = 2 \pi ( m \iota - N), $$
where $N$ is the integer closest to $m \iota$, and if $\iota$ is irrational these values are evenly distributed on the interval $\theta_m \in [-\pi, \pi]$, according to a theorem of Hermann Weyl [\cite{Weyl-1916,Helander-2014-a}]. If $\theta_m$ is small, the local variation of the field strength along the field line is 
	$$  B(\theta, \varphi) \simeq B_{\rm max} + \frac{1}{2} \left[ B_{\theta \theta} (\theta_m + \iota \varphi)^2 + 
	2 B_{\theta \varphi} (\theta_m + \iota \varphi) \varphi + B_{\varphi \varphi} \varphi^2 \right], $$
and its local maximum, which is situated where $\p B/ \p \varphi = 0$, is approximately
	$$ B_{\rm max}^{\rm loc} \simeq B_{\rm max}\left( 1 - \Lambda \theta_m^2 \right), $$
with
	$$ \Lambda B_{\rm max} = \frac{B_{\theta \varphi}^2 - B_{\theta \theta} B_{\varphi \varphi}}{2 ( \iota^2 B_{\theta \theta} + 2 \iota B_{\theta \varphi} + B_{\varphi \varphi})}. $$
The probability that such a local maximum exceeds $B_{\rm max} - \Delta B$ is thus equal to 
	$$ P(B_{\rm max}^{\rm loc} > B_{\rm max} - \Delta B) = \frac{1}{\pi} \sqrt{\frac{\Delta B}{\Lambda B_{\rm max}}}. $$
On average, one thus expects having to follow the field line a number $N(\Delta B) = \pi \sqrt{\Lambda B_{\rm max}/\Delta B}$ turns around the torus, corresponding to a distance
	\bn L(\Delta B) \simeq 2 \pi^2 R \sqrt{\frac{\Lambda B_{\rm max}}{\Delta B}},
	\label{L}
	\en
until $B$ comes within $\Delta B$ of its maximum value.

If collisional scattering changes $\xi$ by $\Delta \xi$, the value of $B$ at the bounce point will change by $\Delta B/B \sim \Delta \xi$. 
We obtain an estimate for the width of the collisional boundary layer by equating the corresponding mean free path (\ref{mfp}) to the length (\ref{L}), which gives
	$$ \Delta \xi \sim \Lambda^{1/5} \nu_\ast^{2/5}, $$
within a factor of order unity. The parameter $\Lambda$ vanishes for omnigeneous stellarators since the maximum field strength on each magnetic surface is not reached in a discrete number of points but along a curve that closes poloidally, toroidally or helically around the torus [\cite{Cary-1997,Helander-2014-a,Parra-2015}]. For optimized stellarators with small $\Lambda$, the estimate $\Delta \xi \sim \Lambda^{1/5} \nu_\ast^{2/5}$ is only valid as long as $L \gg R$, that is, when $\Delta \xi \sim \Delta B/B \ll \Lambda$. For $\Delta \xi \gtrsim \Lambda$, $L \sim R$ and we find a boundary layer width $\Delta \xi \sim \nu_\ast^{1/2}$. Then,
$$ \Delta \xi \sim \left \{ \begin{array}{l l l}
\nu_\ast^{1/2} & \mathrm{for} & \nu_\ast \gtrsim \Lambda^2 \\
\Lambda^{1/5} \nu_\ast^{2/5} & \mathrm{for} & \nu_\ast \lesssim \Lambda^2
\end{array} \right.$$
We thus expect the collisional boundary layer to scale slightly differently in a stellarator as compared with the tokamak. 

It is difficult to say whether this prediction is borne out by numerical simulations. In many cases, the approach to the low-collisionality asymptote given by Eq.~(\ref{bootstrap2}) is so slow that it is impossible to say anything about the scaling, see e.g. Figs.~22-24 of \cite{Beidler-2011} or Fig.~3 of \cite{Kernbichler-2016}. At least sometimes, the finite-collisionality correction appears to scale as $\nu_\ast^{0.5}$, which is indistinguishable from $\nu_\ast^{0.4}$, see \cite{Helander-2011-a}.

\section{Conclusions}

In the present paper, the neoclassical theory of parallel transport in stellarators at low collisionality has been reconsidered. For the $1/\nu$-regime, we have recovered a well-known expression (\ref{Vi}) for the ion flow (plasma rotation) and derived a novel explicit formula (\ref{bootstrap}) for the bootstrap current. Both these results, in particular the former, are however found to be sensitive to the influence of tangential drifts and finite collisionality. Tangential drifts are found to affect the parallel ion flow significantly in the $\sqrt{\nu}$-regime, because they reduce the radial excursion of trapped particle orbits from flux surfaces, and the response of the trapped-particle population to the radial pressure and temperature gradients is therefore weakened. The average flow velocity of the trapped particles is thus affected by the electric field, and this change is transmitted to the passing population by collisions. As a result, the plasma rotation velocity changes significantly.

A finite collision frequency affects the boundary layer between the trapped and passing particles slightly differently from that in a tokamak. The width of this boundary layer is set by the requirement that the effective collisional scattering frequency across the layer should be comparable to the bounce frequency between two local magnetic maxima along the same magnetic field line that are high enough to matter for the boundary layer. Whether they do so depends on the width of this layer, and therefore on the collisionality. In a stellarator, the distance along the magnetic field between two relevant field-strength maxima therefore depends on the collision frequency. As a result, the width of the boundary layer is predicted to scale as $\nu_\ast^{2/5}$ rather than $\nu_{\ast}^{1/2}$.

\section*{Acknowledgment}

The first author gratefully acknowledges the splendid hospitality of Oxford University, in particular Merton College, where this work was initiated. We are grateful to Craig Beidler for many interesting discussions and a number of helpful comments on the manuscript. SLN was supported by the Framework grant for Strategic Energy Research (Dnr. 2014-5392) from Vetenskapsr{\aa}det.

\newpage
\appendix

\section{Model collision operator for electron-electron collisions} \label{app:Cee}

To calculate the electron distribution function, we need a model linearised electron-electron collision operator $C_{ee}(f_e^-)$ that contains the exact pitch-angle scattering piece of the Landau electron-electron collision operator, that conserves momentum, that is self-adjoint, that has $f_e^- = (m_e v_\| V_{e\|}/T_e) F_{e0}$ as the only solution to $C_{ee} (f_e^-) = 0$, and that satisfies (\ref{Sonine condition}). 

Following \cite{Taguchi-1988} and \cite{Helander-2002-c}, we propose the model collision operator
	$$C_{ee} (f_e^-) = \nu_D^{ee} \mathcal{L} ( f_e^- ) + \frac{\nu_D^{ee} m_e v_\|}{T_e} F_{e0} \sum_{p = 0}^S a_p L_p^{(3/2)}(x^2),$$
with the coefficients $a_r$ determined by the system of equations
	\bn \sum_{q = 0}^S N_{pq} a_q = \frac{\tau_{ee}}{n_e} \int \nu_D^{ee} L_p^{(3/2)} (x^2) v_\| f_e^-\, d^3 v.
	\label{Eq for a general}
	\en
The matrix $N_{pq}$ is determined using conditions (\ref{Sonine condition}). For our model collision operator, the left side of Eq.\,(\ref{Sonine condition}) becomes
	\bea\int L_p^{(3/2)} v_\| C_{ee} \left ( L_q^{(3/2)} v_\| F_{e0} \right )\, d^3 v = - \frac{n_e T_e}{m_e} \{ \nu_D^{ee} L_p^{(3/2)} L_q^{(3/2)} \} \\+ n_e \sum_{r = 0}^S a_r \{ \nu_D^{ee} L_p^{(3/2)} L_r^{(3/2)} \}.
	\eea
Using Eq.\,(\ref{Eq for a general}), we find
	\bea\int L_p^{(3/2)} v_\| C_{ee} \left ( L_q^{(3/2)} v_\| F_{e0} \right )\, d^3 v = - \frac{n_e T_e}{m_e} \{ \nu_D^{ee} L_p^{(3/2)} L_q^{(3/2)} \} \\+ \frac{n_e T_e \tau_{ee}}{m_e} \sum_{r = 0}^S \sum_{s = 0}^S \{ \nu_D^{ee} L_p^{(3/2)} L_r^{(3/2)} \} \{ \nu_D^{ee} L_q^{(3/2)} L_s^{(3/2)} \} (N^{-1})_{rs},
	\eea
where $(N^{-1})_{rs}$ is the inverse of the matrix $N_{pq}$. Substituting this result into Eq.\,(\ref{Sonine condition}), we find a linear system of equations that determines $(N^{-1})_{rs}$ and hence $N_{pq}$,
	\bn
	\tau_{ee}^2 \sum_{r = 0}^S \sum_{s=0}^S \{ \nu_D^{ee} L_p^{(3/2)} L_r^{(3/2)} \} \{ \nu_D^{ee} L_q^{(3/2)} L_s^{(3/2)} \} (N^{-1})_{rs} = - K_{pq} + \tau_{ee} \{ \nu_D^{ee} L_p^{(3/2)} L_q^{(3/2)} \} .
	\label{invNrs}
	\en
Note that this equation gives a symmetric matrix $N_{pq}= N_{qp}$. We have used Eq.\,(\ref{invNrs}) and
	\bea
	\{ \nu_D^{ee} \} = \frac{0.533}{\tau_{ee}}, \quad \{ \nu_D^{ee} L_1^{(3/2)} \} = \frac{0.625}{\tau_{ee}}, \quad \{ \nu_D^{ee} L_2^{(3/2)} \} = \frac{0.652}{\tau_{ee}}, \\ 
	\{ \nu_D^{ee} (L_1^{(3/2)})^2 \} = \frac{1.386}{\tau_{ee}}, \quad \{ \nu_D^{ee} L_1^{(3/2)} L_2^{(3/2)} \} = \frac{1.563}{\tau_{ee}}, \quad \{ \nu_D^{ee} (L_2^{(3/2)})^2 \} = \frac{2.484}{\tau_{ee}},
	\eea
to determine the coefficients of Eq.\,(\ref{Eq for a}).

The collision operator has been completely determined using condition (\ref{Sonine condition}). It is trivial to check that it contains the pitch-angle scattering piece of the Landau collision operator. We proceed to show that it satisfies the other properties postulated above.

\subsection{Self-adjointness}
The self-adjointness of the collision operator is derived from the fact that $N_{pq}$ is symmetric. Indeed,
	$$\int \frac{g_e^-}{F_{e0}} C_{ee} ( f_e^- )\, d^3 v = \int \nu_D^{ee} \frac{g_e^-}{F_{e0}} \mathcal{L} \left ( f_e^- \right )\, d^3 v + \frac{m_e}{T_e} \sum_{p = 0}^S a_p \int \nu_D^{ee} L_p^{(3/2)} v_\| g_e^-\, d^3 v.$$
Using (\ref{Eq for a general}), we can rewrite this equation as
	\bea\int \frac{g_e^-}{F_{e0}} C_{ee} ( f_e^- )\, d^3 v = \int \nu_D^{ee} \frac{g_e^-}{F_{e0}} \mathcal{L} \left ( f_e^- \right )\, d^3 v \\+ \frac{\tau_{ee} m_e}{ n_e T_e} \sum_{p=0}^S \sum_{q = 0}^S (N^{-1})_{pq} \int \nu_D^{ee} L_p^{(3/2)} v_\| f_e^-\, d^3 v\, \int \nu_D^{ee} L_q^{(3/2)} v_\| g_e^-\, d^3 v,
	\eea
proving self-adjointness. 

\subsection{Conservation of momentum}
The coefficients $N_{pq}$ ensure conservation of momentum. Using the property $K_{p0} = K_{0p} = 0$ in Eq.\,(\ref{invNrs}) gives
	$$\sum_{r = 0}^S \{ \nu_D^{ee} L_p^{(3/2)} L_r^{(3/2)} \} \left ( \sum_{s = 0}^S  (N^{-1})_{rs} \tau_{ee} \{ \nu_D^{ee} L_s^{(3/2)} \} \right )  = \{ \nu_D^{ee} L_p^{(3/2)} \}.$$
This is a linear system of equations for $\sum_{s=0}^S  (N^{-1})_{rs} \tau_{ee} \{ \nu_D^{ee} L_s^{(3/2)} \}$ whose solution is
	$$\sum_{s=0}^S  (N^{-1})_{rs} \tau_{ee} \{ \nu_D^{ee} L_s^{(3/2)} \} = \delta_{0r},$$
where $\delta_{pq}$ is the Kronecker delta. Multiplying this equation by $N_{pr}$ and summing over $r$ finally gives
	\bn N_{p0} = \tau_{ee} \{ \nu_D^{ee} L_p^{(3/2)} \}.
	\label{Np0}
	\en
With this result, we can prove conservation of momentum. The parallel velocity moment of the collision operator is
	$$\int v_\| C_{ee} \left ( f_e^- \right )\, d^3 v = - \int \nu_D^{ee} v_\| f_e^-\, d^3 v + n_e \sum_{p= 0}^S a_p \{ \nu_D^{ee} L_p^{(3/2)} \}.$$
or, using Eq.\,(\ref{Eq for a general}),
	$$\int v_\| C_{ee} \left ( f_e^- \right )\, d^3 v = n_e \sum_{p=0}^S a_p \left ( \{ \nu_D^{ee} L_p^{(3/2)} \} - \frac{N_{p0}}{\tau_{ee}} \right ).$$
Since the coefficients $N_{p0}$ have the property (\ref{Np0}), the collision operator conserves momentum.

\subsection{Solution to $C_{ee}(f_e^-) = 0$}
The solution to $C_{ee} (f_e^-) = 0$ is
\begin{equation}
f_e^- = \frac{m_e v_\|}{T_e} F_{e0} \sum_{p=0}^S a_p L_p^{(3/2)} (x^2)
\end{equation}
(recall that $f_e^-$ is odd in $v_\|$ and that it does not depend on gyrophase). We determine the coefficients $a_p$ by using Eq.\,(\ref{Eq for a general}). For this particular form of $f_e^-$, Eq.\,(\ref{Eq for a general}) becomes
\begin{equation}
\sum_{q = 0}^S \left [ N_{pq} - \tau_{ee} \{ \nu_D^{ee} L_p^{(3/2)} L_q^{(3/2)} \} \right ]a_q = 0.
\end{equation}
This system of equations implies that $a_p = 0$ for $p \neq 0$, and that $a_0$ is a free parameter since the coefficients $N_{p0}$ satisfy condition (\ref{Np0}). Hence $f_e^- = (m_e v_\| a_0/T_e) F_{e0}$ is the only solution to $C_{ee} (f_e^-) = 0$.


\bibliographystyle{jpp}
\bibliography{pdh}

\end{document}